%% file: main.tex
\documentclass[10pt,conference]{IEEEtran}
\IEEEoverridecommandlockouts
\makeatletter
\newcommand*\titleheader[1]{\gdef\@titleheader{#1}}
\AtBeginDocument{%
  \let\st@red@title\@title
  \def\@title{%
    \bgroup\normalfont\large\centering\@titleheader\par\egroup
    \vskip1.5em\st@red@title}
}
\makeatother

\usepackage{stackengine}
\usepackage{cite}
\usepackage{subfig}
\usepackage{amsmath,amssymb,amsfonts}
\usepackage{algorithmic}
\usepackage{graphicx}
\usepackage{textcomp}
\usepackage{xcolor}
\usepackage[ruled,linesnumbered]{algorithm2e}
\usepackage[utf8]{inputenc}
\usepackage{enumitem}
\usepackage{tabularx,multirow,multicol,booktabs,colortbl}
\newcommand{\code}[1]{\texttt{#1}}
\newcommand{\nocbase}{NOC\_BASE}
\newcommand{\nocsw}{NOC\_SW}
\newcommand{\nocswc}{NOC\_SW\_C}
\newcommand{\axis}{AXI4-Stream}
\newcommand{\exrocket}{\code{ExampleRocketSystem}}
\newcommand{\node}[1]{$N_{#1}$}

\newcommand{\nocfunc}[1]{\code{noc\_#1()}}
\def\BibTeX{{\rm B\kern-.05em{\sc i\kern-.025em b}\kern-.08em
    T\kern-.1667em\lower.7ex\hbox{E}\kern-.125emX}}

\title{ANDROMEDA: An FPGA Based RISC-V MPSoC Exploration Framework\vspace{-3mm}}
\titleheader{Preprint - Accepted in VLSI Design 2021\vspace{-15mm}}
\author{\IEEEauthorblockN{Farhad Merchant, Dominik Sisejkovic, Lennart M. Reimann, \\ Kirthihan Yasotharan, Thomas Grass, Rainer Leupers} 
		\IEEEauthorblockA{Institute for Communication Technologies and Embedded Systems, RWTH Aachen University, Germany} 
        \{farhad.merchant, sisejkovic, reimannl, yasotharan, grass, leupers\}@ice.rwth-aachen.de \vspace{-7mm}}

\begin{document}
\bstctlcite{IEEEexample:BSTcontrol} 

\maketitle

\begin{abstract}
\input{abstract}
\end{abstract}


\section{Introduction}
\input{introduction}

\section{Background and Related Work}\label{sec:rw}
\input{rw}

\section{Network-on-Chip and Andromeda}\label{sec:noc}
\input{noc}


\section{Experimental Setup and Results}\label{sec:exp}
\input{results}

\section{Conclusion}\label{sec:con}
\input{conclusion}

\bibliographystyle{IEEEtran}
\bibliography{IEEEabrv,ref}

\end{document}

%% file: abstract.tex
With the growing demands of consumer electronic products, the computational requirements are increasing exponentially. Due to the applications' computational needs, the computer architects are trying to pack as many cores as possible on a single die for accelerated execution of the application program codes. In a multiprocessor system-on-chip (MPSoC), striking a balance among the number of cores, memory subsystems, and network-on-chip parameters is essential to attain the desired performance. In this paper, we present \emph{ANDROMEDA}, a RISC-V based framework that allows us to explore the different configurations of an MPSoC and observe the performance penalties and gains. We emulate the various configurations of MPSoC on the Synopsys HAPS-80D Dual FPGA platform. Using STREAM, matrix multiply, and N-body simulations as benchmarks, we demonstrate our framework's efficacy in quickly identifying the right parameters for efficient execution of these benchmarks. 

%% file: introduction.tex
Multiprocessor system-on-chips (MPSoCs) are the common computing substrate for application domains such as automotive, and internet-of-things (IoT)~\cite{auto1}\cite{iot1}. Recently, MPSoCs are heavily involved in service-oriented architectures due to their capabilities to offer an order of speed-up over the state-of-the-art~\cite{sosoc}.  The applications executed on the MPSoCs in these domains are compute and communication-intensive, requiring complex platforms with deep memory hierarchy and network infrastructure for reduced application run-time at constrained energy and area footprints~\cite{ref1}. A right balance between system parameters is required to attain the desired performance. 

Exploring system-level parameters for run-time performance improvement is an extensively studied topic in the literature~\cite{esl1}. However, most of the literature solutions are incomplete and sometimes rely on a complex set of parameters that can be simplified. Also, for many solutions, the rapid field-programmable gate array (FPGA) prototyping is intricate due to complexities involved in the intermediate tools, since the tools have non-standardized tool-interfaces.   

Recently, an open instruction-set architecture called RISC-V has revolutionized the system design aspects due to its flexibility~\cite{riscv_priv}. Due to the momentum gained by the adoption of RISC-V, there is an increasing demand for RISC-V based system design and prototyping. While there have been several attempts to design and develop efficient single-core RISC-V compute platforms, there are a handful of MPSoCs~\cite{mpsoc1}\cite{ariane1}. There is an increasing need for the frameworks to support RISC-V based MPSoC exploration to enable the system designers to develop efficient platforms for the next-generation computing systems.

In this paper, we present \emph{ANDROMEDA}, a unified framework that facilitates the design space exploration of FPGA based MPSoCs. The ANDROMEDA framework helps in identifying the bottlenecks in application execution for system-level optimizations. The major contributions in this paper are as follows:

\begin{itemize}
    \item A light-weight network-on-chip (NoC) for clustered RISC-V MPSoC platform development.
    \item ANDROMEDA, an FPGA-based MPSoC framework for early-stage exploration and application execution bottleneck identification.
    \item Evaluation of ANDROMEDA using STREAM, matrix multiply, and N-body simulations, and identification of bottlenecks for the benchmarks.
\end{itemize}

The rest of the paper is organised as follows: In Section~\ref{sec:rw}, we discuss background and the literature. A light-weight NoC implementation is discussed in Section~\ref{sec:noc} along with the proposed ANDROMEDA framework. The experimental setup and results are depicted in Section~\ref{sec:exp}. We conclude our work in Section~\ref{sec:con}. 

%% file: rw.tex
\subsection{Background}
\subsubsection{RocketChip Generator}
The Rocket Chip is an open-source tool to instantiate the Rocket Core, and synthesizable system-on-chip (SoC). The Rocket Chip generator tool is implemented in the \emph{Chisel} hardware construction language. The major advantage of the RocketChip generator is its configurability. The parameters such as, number of cores, cache size, cache placement policies, arithmetic units, pipeline stages, memory management unit, hardware performance counters, interconnect, and others can be customised to attain the SoC.

\subsubsection{Rocket core}
Rocket is a 5-stage single-issue in-order CPU which implements the RISC-V ISA. It supports both RV32G and RV64G. Rocket features a non-blocking data cache, a branch predictor, an MMU, and an FPU. Rocket can be configured to meet individual requirements. Options include supported ISA extensions (e.g., M, A, F, D) and cache sizes.


\subsubsection{Synopsys HAPS-80D Dual}

\begin{figure}[!t]
\centering
\includegraphics[width=0.85\columnwidth]{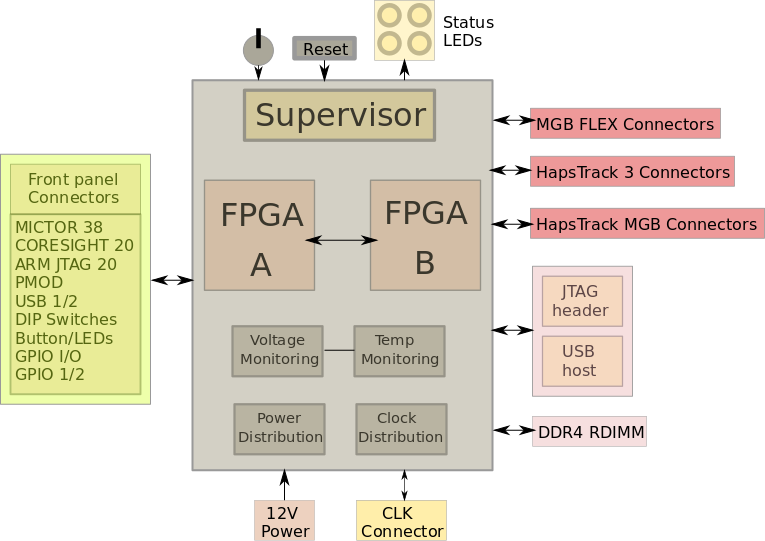}
\caption{Synopsys HAPS-80D Dual}
\label{fig:haps}
\end{figure}

Synopsys high-performance ASIC prototyping systems (HAPS) is a family of FPGA based systems for prototyping ASICs and SoCs. HAPS solutions are shipped with prototyping hardware and the supporting software tools, to enable faster system validation and earlier software development. 

Fig.~\ref{fig:haps} shows the overall system architecture of HAPS-80D Dual platform. The platform consists of two UltraScale XCVU440 FPGAs. The FPGAs are connected via several high-speed serial links. HAPS-80D Dual is equipped with 13 low-skew clock networks. The global clocks are labeled from \texttt{GCLK0} to \texttt{GCLK2}. \texttt{GCLK0} is a fixed 100 MHz clock reserved for system function. \texttt{GCLK1}-\texttt{GCLK12} are available for user designs, and in most cases, are sourced from the on-board phase-lock-loops (PLLs). The HAPS-80D Dual contains an on-board DDR4 DRAM module with 8GB capacity. It can be used as memory in user designs, or as a large sample memory for signal debugging. HAPS-80D Dual features various I/O options, such as GPIO, PMOD, JTAG and UART. The HAPS-80D Dual system connects to a host computer via a high-speed USB-C cable. The Synopsys UMRBus protocol is used for both configuring the system with the user design, and for subsequent debugging. 

\subsection{Related Work}

There has been a plethora of MPSoC works in the literature focusing on early stage design space exploration~\cite{esl1}. Some of the early works focused on design automation for custom generation of MPSoCs, while a few of them focused on industrial-grade design customization. Daedalus framework presented in~\cite{daedalus1} and~\cite{daedalus2} introduces a system-level exploration, programming and prototyping framework. The Daedalus framework takes a sequential application as an input and translates it into an MPSoC implementation on FPGA. The Daedalus framework considers only dataflow dominated applications. 

The system-on-chip environment (SCE) is an interactive framework presented in~\cite{sce1}. The SCE frameworks accepts high-level specifications as an input and translates the specifications into hardware/software implementation. The major bottleneck in the SCE design flow is the initial system-level specifications for the desired hardware/software implementation. The SystemCoDesigner framework presented in~\cite{systemcodesigner1} has similar limitations. 

Our proposed framework, ANDROMEDA, is an over-simplistic framework for RISC-V based MPSoC prototyping on FPGA. The input to the framework are several high-level system parameters that result in an MPSoC implementation on Synopsys HAPS-80D Dual platform.

%% file: noc.tex
\begin{table}[!b]
	\centering
	\tiny
	\caption{NoC configurations\vspace{-2mm}}
	\begin{tabular}{| c | c | c | c | c |}
		\hline
		\textbf{Name} & \textbf{Nodes} & \textbf{Proc. cores / node} & \textbf{Router} & \textbf{Flow-control} \\
		\hline
		\nocbase & 16 & 4 & $S_k$ & Store-and-forward \\
		\nocsw & 16 & 4 & AXIS-Switch & Store-and-forward \\
		\nocswc & 16 & 4 & AXIS-Switch & Cut-through \\
		\hline
        
	\end{tabular}
	\label{tab:noc_configs}
\end{table}
A base system, \nocbase, was designed, which was used as a starting point for design space exploration. Table~{\ref{tab:noc_configs}} shows the parameters of \nocbase, together with other configurations, which are explored in later section. In the following, the hardware architecture of \nocbase\ is covered.
\begin{figure}[!t]
	\centering
	\includegraphics[width=0.55\columnwidth]{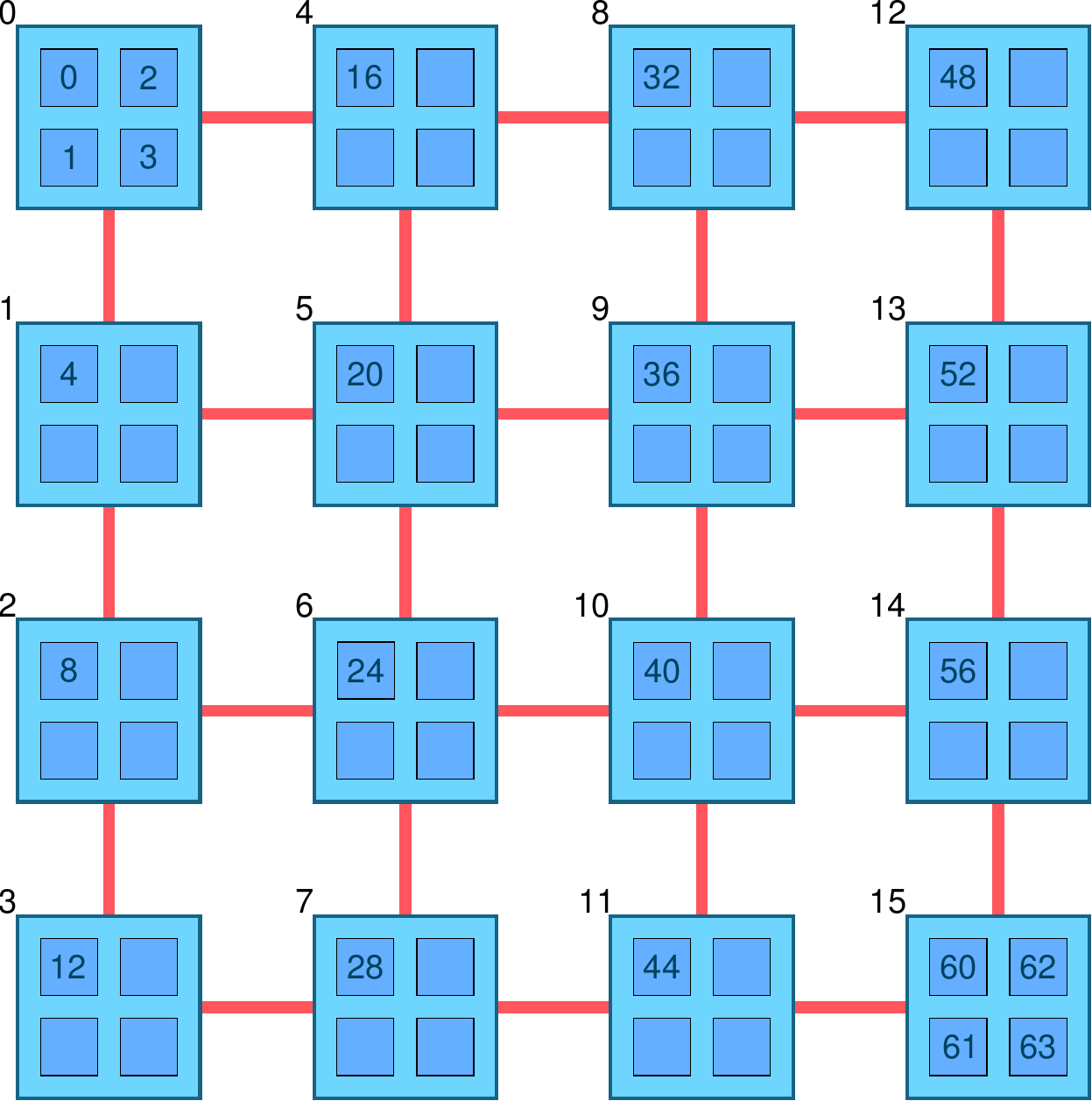}
	\caption{The RocketChip mesh NoC: numbers at top left corners of nodes indicate \textit{node IDs}; numbers inside nodes indicate \textit{core (hart) IDs}.}
	\label{fig:rocket_noc}
\end{figure}

\begin{figure}[!t]
	\centering
	\includegraphics[width=0.6\columnwidth]{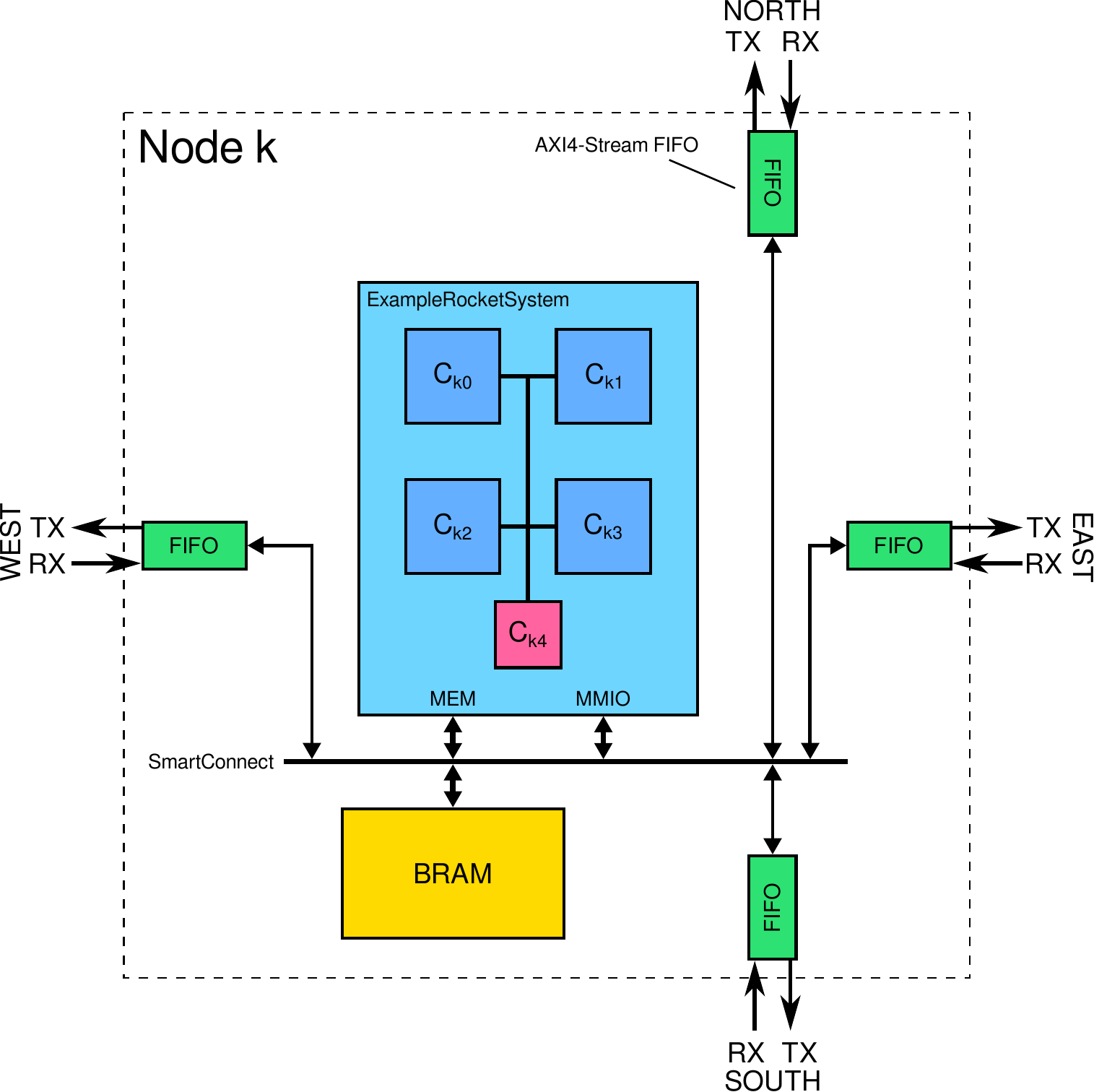}
	\caption{Internal structure of a node in NOC\_BASE}
	\label{fig:rocket_node}
\end{figure}


The system consists of 16 nodes, $N_0 ... N_{15}$, interconnected in a $4 \times 4$ 2-D mesh network, as shown in Fig.~{\ref{fig:rocket_noc}}. Inside each node \node{k}, a RocketChip SoC with  $(n+1)$ cores, $C_{k0} ... C_{kn}$ (hart 0 ... hart $n$) is located. Every RocketChip instance is created from the same generated \exrocket. Cores $C_{k0} ... C_{k,(n-1)}$ are (general-purpose) processing cores, while $C_{kn} =: S_k$ is a \textit{small} RocketCore, which is also coherently interconnected with the processing cores. The small core's role is to act as a co-processor to handle all network-related tasks, like routing and switching. Though using a general-purpose core as a software router is likely not the optimal solution, this route was initially chosen to get to a working prototype system as quickly as possible. Furthermore, without the RocketChip generator, the design of this particular approach would have been more involved, too. 

\begin{table}[!b]
    \tiny
	\centering
	\caption{Data cache configurations\vspace{-2mm}}
	\begin{tabular}{| c | c | c | c |}
		\hline
		\textbf{Name} & \textbf{nSets} & \textbf{nWays}& \textbf{Size (KiB)} \\
		\hline
		\textit{BASE} & \textit{64} & \textit{4} & \textit{16} \\
		C-64-8 & 64 & 8 & 32 \\
		C-64-16  & 64 & 16 & 64 \\
		\hline
	\end{tabular}
	\label{tab:expl_cache}
\end{table}

\begin{figure}[!t]
	\centering
	\includegraphics[width=0.95\columnwidth]{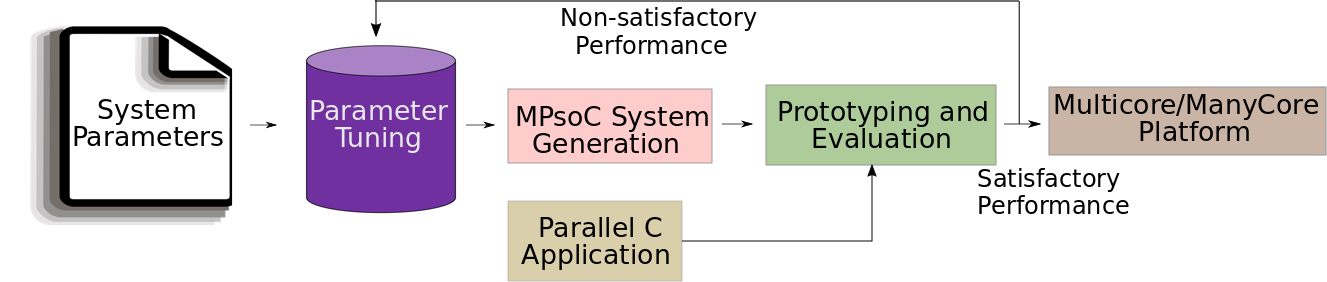}
	\caption{ANDROMEDA framework for FPGA based MPSoC exploration\vspace{-5mm}}
	\label{fig:andromeda}
\end{figure}

The memory model of the system follows the \textit{distributed memory} paradigm. Each node contains BRAM as main memory (used for data and instructions), which is only accessible by the node's local cores, and not from other nodes. The size of the BRAM is configurable in Vivado and was set to be 256 kB for all nodes, except \node{0}, which has 2 MB. In total, the complete system can fit up to 5.75 MB of on-chip data. \node{0} additionally incorporates an AXI UART Lite to interface with a serial console on the host computer. Fig.~{\ref{fig:rocket_node}} shows the internal structure of a node. We explain usage of NoC and RocketChip to incorporate a distributed memory system, ANDROMEDA for system level exploration (see Fig.~\ref{fig:andromeda}). 

ANDROMEDA is a simple framework that consisting of input system parameters that are used for system configuration. The parameters are the number of cores, cache size (see configuration in Table~\ref{tab:expl_cache}), and coherency, floating-point unit (FPU) and its pipeline stages, memory management unit, hardware performance counters, and interconnect network and its parameters. 

In ANDROMEDA, a designer sets the parameters that automatically generate the desired multicore or manycore system. The generated system is later prototyped on the HAPS-80D Dual FPGA prototyping platform, where the system is evaluated using benchmarks written in the C programming language. If the performance attained is not satisfactory, then the parameters are calibrated through the designer intervention to arrive at a satisfactory performance. The proposed ANDROMEDA framework enables rapid prototyping as the system parameters are at a higher level of abstraction than in the literature. At this stage, automatic tuning of the parameters is out of the scope of this work.

%% file: results.tex
\subsection{Benchmarks}
To evaluate the performance of ANDROMEDA, three benchmarks written in C have been applied. In the following, the selected benchmarks are briefly explained. 

\subsubsection{STREAM}

The STREAM benchmark~\cite{streambench1} is used to measure the \textit{sustainable memory bandwidth} of computer systems. Though initially targeted for high-performance computing systems, it can also be applied on personal or embedded computers. STREAM executes four kernels on contiguous arrays \texttt{a}, \texttt{b} and \texttt{c}, and determines from the execution time the resulting bandwidth. The kernels are: COPY (\texttt{a[i] = b[i]}), SCALE (\texttt{a[i] = q*b[i]}), ADD (\texttt{a[i] = b[i] + c[i]}), and TRIAD (\texttt{a[i] = b[i] + q*c[i]}).
    
%
%

To minimize cache effects, the sizes of the arrays should, in general, be larger than the largest available cache. The data type of the arrays can be selected in the benchmark, with the default type being \texttt{double}, which was also used in this work.

\subsubsection{Matrix Multiplication}
As matrix multiplication (Matmul) is an operation found in many applications, a Matmul benchmark was also implemented for RocketChip. It multiplies two matrices, $A\ (N \times K)$ and $B\ (K \times M)$, and writes the result into a third matrix, $C\ (N \times M)$. Matrix $A$ and $C$ are stored in \textit{row-major} format, while matrix $B$ is stored \textit{column-major}. All matrices contain \texttt{double-precision} values.

\subsubsection{N-body Simulation}
The N-body simulation (N-body) numerically solves the \textit{N-body problem}, which is a classic problem from orbital mechanics. Given are $N$ bodies with mass $m_i$, initial position $\vec{r_i}(t=0)$ and initial velocity $\vec{v_i}(t=0)$. The bodies exert forces on each other according to Newton's gravitation law. The net force exerted on body $i$ is then given by:

\begin{equation}\label{eq:newton}
    \vec{F_i}(t) = \sum_{\substack{j=0 \\ j \neq i}}^{N-1}{\frac{G m_i m_j}{\left\lVert \vec{r_i}(t) - \vec{r_j}(t) \right\rVert^3} (\vec{r_i}(t) - \vec{r_j}(t))}.
\end{equation}

The goal is to find the positions and velocities of each of the bodies after a time $t$. The simulation is performed in several \textit{timesteps}. In each timestep, the net force exerted on each body is calculated according to equation \ref{eq:newton}. From the net force, the acceleration experienced by that particular body is computed. Then, based on the acceleration, the new position and velocity can be derived. The computational complexity of an N-body simulation is $\mathcal{O}(N^2)$. All values (masses, positions, velocities) are stored in single-precision floating-point format (\texttt{float}). The storage requirement scales with $\mathcal{O}(N)$.

%


        
        

\subsection{Parallelization Techniques}

For parallelizing the benchmarks, the \textit{worksharing} principle was applied. In all three benchmarks (STREAM, Matmul, N-body), the work can be distributed evenly among the cores. For bare-metal applications, some manual work is required to realize the worksharing. In general, each core calculates the start index of its designated partition based on its \textit{hart ID}. After a parallel region, synchronization may be necessary. The \texttt{riscv-tests} provide a \texttt{barrier()} function, which can be used to synchronize the cores.

Matmul can be parallelized by dividing matrix $A$ into sub-matrices $A_{i}$. Each core $i$ then calculates the corresponding partition $C_{i}$ of matrix $C$ by multiplying  $A_{i}$ with $B$.

N-body is parallelized by partitioning the arrays containing the body data evenly among the cores. Each core then computes the new positions and velocities of the bodies in its assigned partition. At the end of each timestep, a barrier is required, to ensure that each core has updated its body values before the next timestep starts.

\subsection{Evaluation}

This section evaluates the ANDROMEDA presented in section~\ref{sec:noc} for different parameters and evaluates single node (BASE) resource utilization and benchmark performance. BASE32 represents single node with 32 cores. 

\subsubsection{Number of cores}

\paragraph{FPGA utilization}

Table~{\ref{tab:eval_multifpgautil}} shows the FPGA resource utilization of BASE for core counts from one to 32. Synthesis for all versions was constrained to 10 MHz. As can be seen in table~{\ref{tab:eval_multifpgautil}}, the LUT counts seem to suggest sub-linear scaling. However, this is primarily caused by modules inside the SoC that are independent of the number of cores. A linear increase can be observed for the number of BRAM, which are used for the L1 data and instruction caches, and DSP slices, which are, for example, used to implement parts of the floating-point unit (FPU). The 32 core system utilizes around 40\% of the available LUTs.

\begin{table}[!b]
    \tiny
	\centering
	\caption{FPGA utilization of BASE (f = 10 MHz)\vspace{-2mm}}
	\begin{tabular}{| c | c | c | c |}
		\hline
		\textbf{Number of cores} & \textbf{LUTs} & \textbf{BRAM} & \textbf{DSP} \\
		\hline
		\hline
		1   & 37,324      ($\times1$)     & 12 & 35 \\
		2   & 69,053      ($\times1.9$)   & 24 & 70 \\
		4   & 131,188     ($\times3.5$)   & 48 & 140 \\
		8   & 255,000     ($\times6.8$)   & 96 & 280 \\
		16  & 499,931    ($\times13.4$)  & 192 & 560 \\
		32  & 998,145    ($\times26.7$)  & 384 & 1120 \\
		\hline
	\end{tabular}
	\label{tab:eval_multifpgautil}
\end{table}

\begin{figure*}[!t]
	\centering
	\subfloat[]{
        \includegraphics[width=0.66\columnwidth]{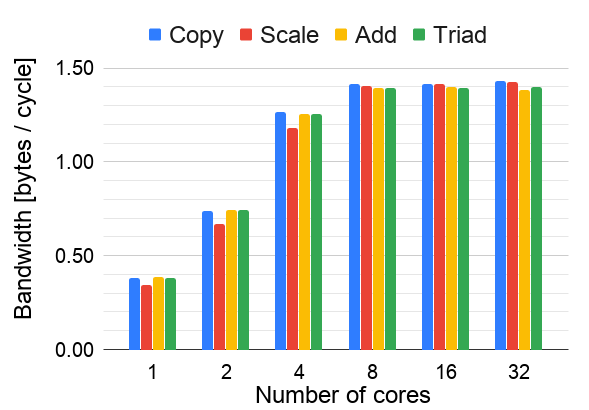}
        \label{fig:stream_base}}
    \subfloat[]{
        \includegraphics[width=0.66\columnwidth]{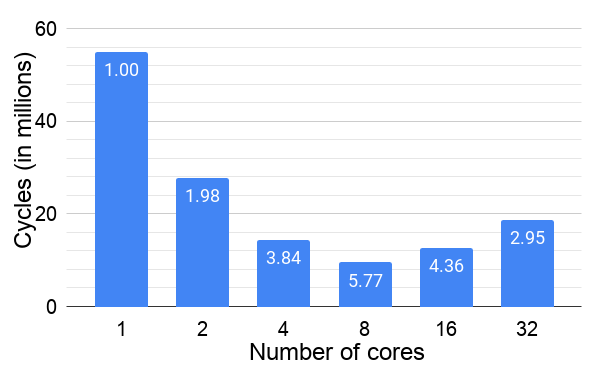}
        \label{fig:matmul_base32}}
    \subfloat[]{
        \includegraphics[width=0.66\columnwidth]{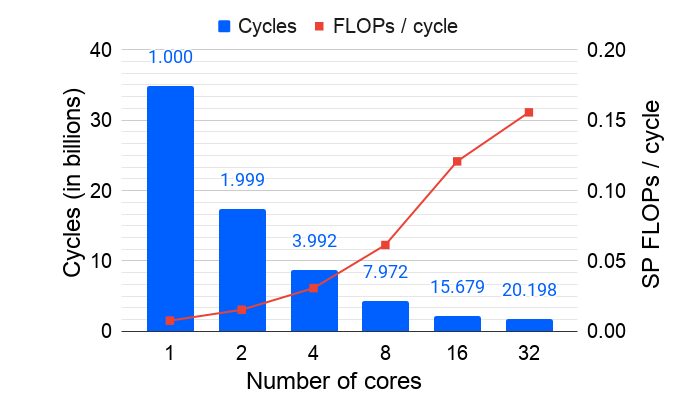}
        \label{fig:multi_nbody}}
 	\caption{(a) STREAM benchmark run on BASE32, size = 128,000 doubles, (b) Matmul (N=K=M=128) run on BASE32, and (c) N-body simulation with N=4,096 bodies and 10 timesteps run on BASE32.}
\end{figure*}

\paragraph{STREAM} Figure~{\ref{fig:stream_base}} shows the memory bandwidth measurements made from the STREAM benchmark. It was run on BASE32 with an array size of 128,000. The theoretical peak memory bandwidth delivered by the outer memory bus (i.e., SmartConnect) is 8 bytes per cycle, as the AXI data bus is 64 bit wide. As Fig.~{\ref{fig:stream_base}} suggests, with increasing core count, a larger fraction of the available memory bus bandwidth could be utilized. A linear scaling can be seen for one to four cores. Beginning at eight cores, however, the sustainable bandwidth begins to saturate at about 1.41 bytes/cycle for the COPY kernel, well below the theoretical maximum. 


\paragraph{Matmul} Matmul with N=K=M=128 was run on BASE32. Fig.~{\ref{fig:matmul_base32}} shows the average execution times (in cycles) for varying core counts. Until four cores, the speedup is approximately equal to the number cores. At eight cores, the speedup drops to about 6 (instead of ideally 8). The situation worsens for even larger core counts. These results again show the limitation of shared memory architectures. 


\paragraph{N-body} Fig.~{\ref{fig:multi_nbody}} plots the execution times (in cycles) of an N-body simulation with N=4,096 bodies run on BASE32. As can be seen from Fig.~{\ref{fig:multi_nbody}}, the cycle count decreases with increasing core count, as expected. Up until 16 cores, close-to-ideal speedup is achieved. For 32 cores, however, the speedup is only 20 (instead of ideally 32). At high core counts, the shared memory bus can still limit the performance. 


Overall it can be concluded that adding more cores, in general, improves performance. However, if the number of cores sharing a bus gets too large ($>$ 8), the limited bus bandwidth caps the performance.

The situation can be improved by reducing the pressure applied on the bus. One method is to increase the size of the caches so that the probability of finding data in the local cache is increased. The next section analyzes the effects of the different cache configurations from Table~{\ref{tab:expl_cache}}.

\subsubsection{Cache subsystem}

\paragraph{FPGA utilization} The synthesis results for the cache configurations from Table~{\ref{tab:expl_cache}} are shown in Table~{\ref{tab:eval_cachefpgautil}}. The percentages indicate the relative increase in the respective resource count compared to BASE. As expected, the usages of both LUTs and BRAMs of the L1 data cache grow as the number of ways is doubled. The last column, "Total LUTs", refers to the LUT usage of the complete \texttt{ExampleRocketSystem} of a particular configuration. It can be seen that the increase of LUTs is reasonable for both cache configurations.

\begin{table}[!b]
    \tiny
    \centering
     \caption{FPGA utilization of cache configs (f = 50 MHz)\vspace{-2mm}}
    \begin{tabular}{|c|c|c|c|}
    \hline
    \multirow{2}{*}{\textbf{Config}} & \multicolumn{2}{c|}{\textbf{L1 Data Cache}}                             & \multicolumn{1}{c|}{\multirow{2}{*}{\textbf{Total LUTs}}} \\ \cline{2-3}
                                     & \multicolumn{1}{c|}{\textbf{LUTs}} & \multicolumn{1}{c|}{\textbf{BRAM}} & \multicolumn{1}{c|}{}                                     \\ \hline
    BASE                             & 3065                               & 4                                  & 131026                                                    \\
    C-64-8                           & 3579 ($+16.8\%$)                              & 8 ($+100\%$)                                  & 133271 ($+1.7\%$)                                           \\
    C-64-16                          & 4248 ($+38.6\%$)                               & 16 ($+300\%$)                                 & 136078 ($+3.9\%$)                                          \\ \hline
    \end{tabular}
    \label{tab:eval_cachefpgautil}
\end{table}

To compare the different L1 data cache sizes, Matmul and N-body were run on the three configurations.

\begin{figure}[!t]
	\centering
	\subfloat[]{
        \includegraphics[width=0.45\columnwidth]{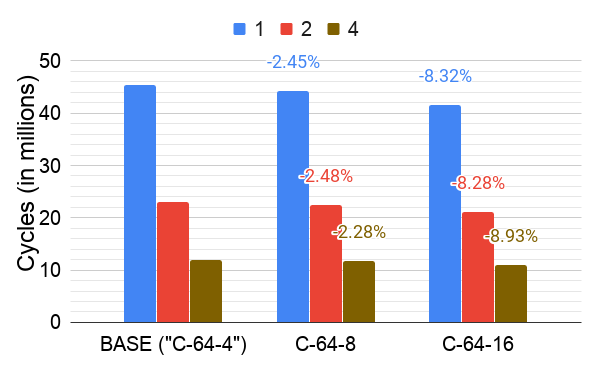}
        \label{fig:cache_matmul}}
    \subfloat[]{
        \includegraphics[width=0.45\columnwidth]{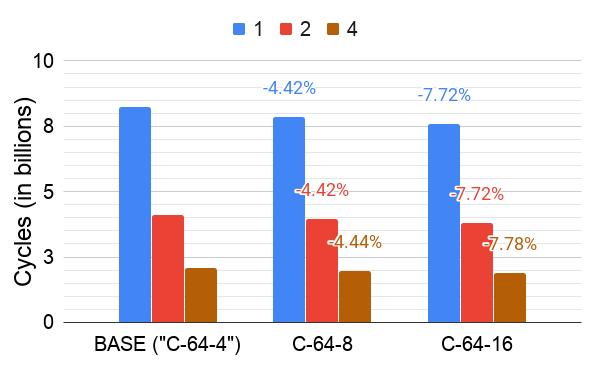}
        \label{fig:cache_nbody}}
    \caption{(a), Matmul (N=K=M=128) run on BASE, C-64-8 and C-64-16 for core counts 1, 2 and 4, and (b) N-body simulation (N=2,048, 10 timesteps) run on BASE, C-64-8 and C-64-16 for core counts 1, 2 and 4.\vspace{-3mm}}
\end{figure}

\paragraph{Matmul} Fig.~{\ref{fig:cache_matmul}} shows the execution time (in cycles) for a Matmul with N=K=M=128 run on the three configurations from Table~{\ref{tab:eval_cachefpgautil}}. It can be seen that in general, the performance improves for larger cache sizes. For a cache size of 64 kb (C-64-16), the execution time drops by over 8\% for all core counts compared to BASE. This result is achieved with an LUT overhead of less than 4\%, but with quadruple the amount of BRAMs per L1 data cache.

\paragraph{N-body} Fig.~{\ref{fig:cache_nbody}} shows the execution time (in cycles) for an N-body simulation with N=2,048 bodies. As for Matmul, it can be observed that the execution time reduces when more cache is available. A larger cache means that more bodies can be stored in the local memories. 


The results show that larger caches indeed improve performance, though depending on the particular application's working set size and access patterns. The total area overhead is mainly devoted to on-chip memory.

\subsection{Network-on-Chip}

The NoC presented in Section~\ref{sec:noc} has been designed to evaluate the benefits of distributing cores across multiple nodes, instead of having the cores share a single bus. This section evaluates the explored NoC configurations from Table~{\ref{tab:noc_configs}}.

\subsubsection{FPGA synthesis results}

Table~{\ref{tab:eval_nocfpgautil_total}} summarizes the total resource usage of the three explored configurations. Both \nocsw\ and \nocswc\ reduce the overall utilization compared to \nocbase. \nocsw\ and \nocswc, both use hardware switches instead of a software router. \nocswc\ uses a switch with cut-through switching.  The primary savings in terms of LUTs and BRAMs are the consequence of removing three \axis\ memory-mapped FIFOs from each node. LUTs can be saved because the \axis\ Data FIFOs, added in \nocsw\ and \nocswc, require less logic as they do not come with an MMIO interface. 
\begin{table}[!b]
    \tiny
    \centering
    \caption{FPGA utilization of NoC configs (f = 5 MHz)\vspace{-2mm}}
        \begin{tabular}{|c|c|c|}
        \hline
        \textbf{Config} & \textbf{LUTs}       & \textbf{BRAM}     \\ \hline
        \nocbase       & 2,011,152           & 2,384         \\
        \nocsw         & 1,905,056 ($-5.3\%$) & 2,368 ($-0.7\%$) \\
        \nocswc      & 1,901,234 ($-5.5\%$) & 2,368 ($-0.7\%$) \\ \hline
        \end{tabular}
        \label{tab:eval_nocfpgautil_total}

\end{table}
\begin{table}[!b]
    \tiny
    \centering
    \caption{FPGA utilization of NoC configs (f = 5 MHz)\vspace{-2mm}}
        \begin{tabular}{|c|c|}
        \hline
        \textbf{Config} & \textbf{LUTs}    \\ \hline
        NOC\_BASE       & 52,992           \\
        NOC\_SW         & 39,016 ($-26.4\%$) \\
        NOC\_SW\_C      & 35,880 ($-32.3\%$) \\ \hline
        \end{tabular}
        \label{tab:eval_nocfpgautil_net}
    
    \label{tab:eval_nocfpgautil}
\end{table}

\subsubsection{Benchmarks}

The performance and scaling of the NoC is analyzed using the previously introduced Matmul and N-body benchmarks. 

\paragraph{Matmul} For parallelizing Matmul on the NoC, the principles from the shared memory parallel version presented earlier were extended to multiple nodes. Now, each node \node{k} is assigned a partition $A_{k}$. The partitions are distributed by \node{0} using the \nocfunc{scatter} function. Further, each node needs the complete matrix $B$ to compute its respective partition $C_k$ of matrix $C$. $B$ is distributed by \node{0} using the \nocfunc{bcast} function. The calculated partitions $C_k$ are finally collected by \node{0} with the \nocfunc{gather} function.

\begin{figure*}[!t]
	\centering
	\subfloat[]{
        \includegraphics[width=0.66\columnwidth]{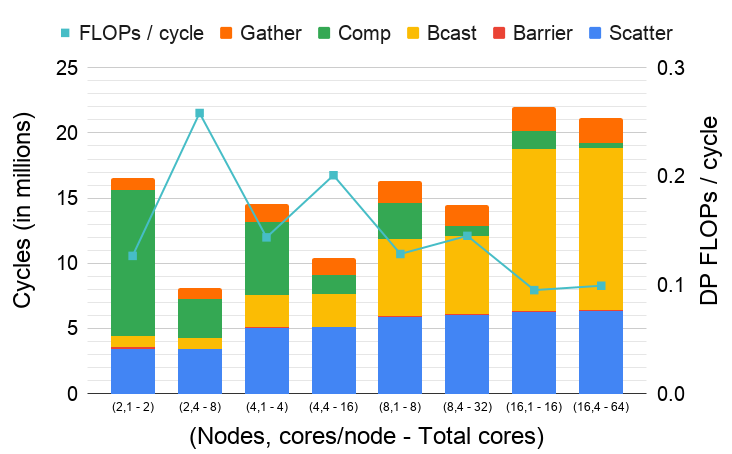}
        \label{fig:noc_matmul2}}
    \subfloat[]{
        \includegraphics[width=0.66\columnwidth]{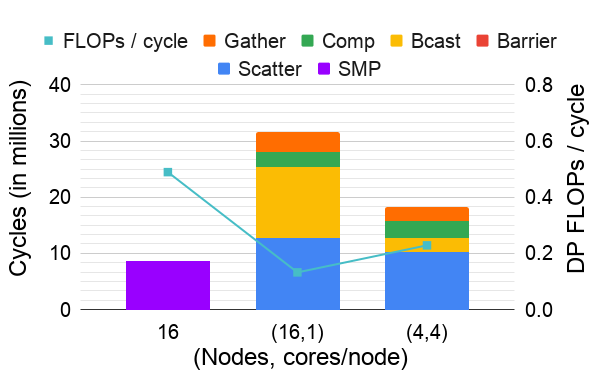}
        \label{fig:noc_matmul3}}
    \subfloat[]{
        \includegraphics[width=0.66\columnwidth]{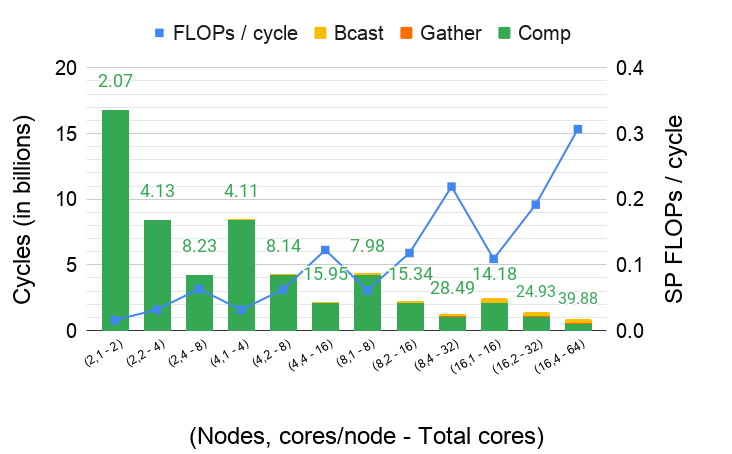}
        \label{fig:noc_nbody2}}
	\caption{(a) Matmul (N=256, K=128, M=32) run on NOC\_SW\_C. (b) Comparison of Matmul (N=512, K=128, M=32) run on NOC\_SW\_C and 16-core BASE32, and (c) N-body simulation (N=4,096) run on NOC\_SW\_C. The values at the top of each bar indicate the speedup compared to the single core version from Fig.~{\ref{fig:multi_nbody}}.}
\end{figure*}

Fig.~{\ref{fig:noc_matmul2}} shows the execution times (in cycles) for a Matmul with N=256, K=128 and M=32, run on different node/core arrangements. It can be observed that the computation time decreases with increasing aggregate core count. This scaling even holds for the maximum core count of 64. Recall, that in the shared memory case, the cycle count began to increase already for core counts above eight. Fig.~{\ref{fig:noc_matmul3}} compares the execution times for a Matmul (N=512, K=128, M=32) between two 16 core arrangements, (16,1) and (4,4), and a 16-core SMP system. It can be seen that even though the 16-core SMP system suffers from the scaling bottleneck as seen in Fig.~{\ref{fig:matmul_base32}}, it still performs better than the two compared node/core configurations, which are dominated by the communication overheads.


In conclusion, the naive parallel Matmul is not well suited for execution on a distributed memory system, where communication dominates computation. Though exploiting \textit{intra-node} parallelism improves performance in general, the gains are quickly overshadowed by the communication overhead, as more nodes are introduced. Lastly, it must be again stated that the potential of the NoC designed in this work is likely limited by the small cores $S_k$.

\paragraph{N-body} As the N-body simulation is a computationally intensive application, it appears to be a good candidate to benefit from a distributed system. The NoC parallel version follows similar principles as the shared memory version. 

Fig.~{\ref{fig:noc_nbody2}} shows the execution times (in cycles) for an N-body simulation with N=4,096. The speedup values compared to the single core version from Fig.~{\ref{fig:multi_nbody}} are indicated at the top of each bar. It can be immediately noticed that the computation to communication ratio is considerably larger than for Matmul. Further, the speedup is \textit{superlinear} up until arrangement (4,2) compared to the speedups of the SMP version shown in Fig.~{\ref{fig:multi_nbody}}. 

\begin{figure}[h!]
	\centering
	\subfloat[16 cores]{
        \includegraphics[width=0.45\columnwidth]{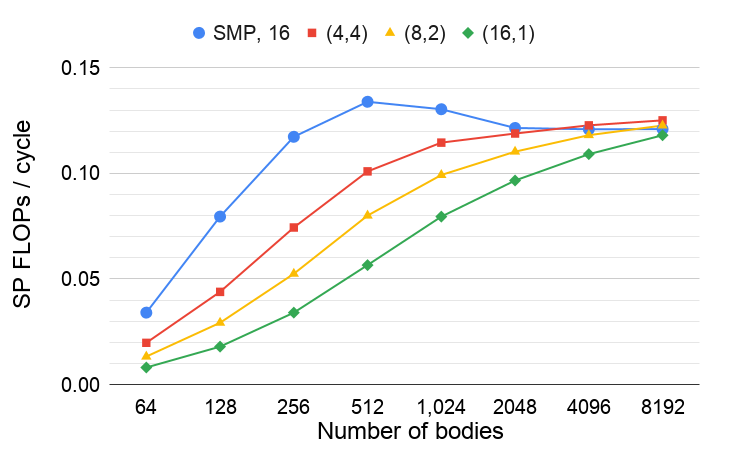}
        \label{fig:noc_nbody3}}
    \subfloat[32 cores]{
        \includegraphics[width=0.45\columnwidth]{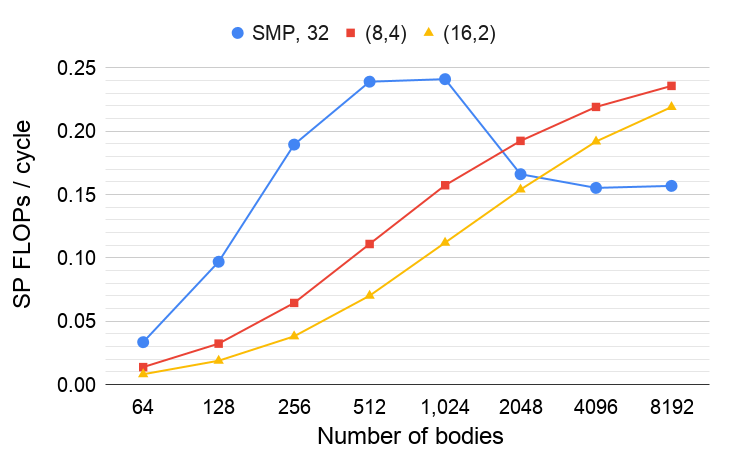}
        \label{fig:noc_nbody4}}
	\caption{Comparison of N-body between NOC\_SW\_C and two SMP systems (BASE32)\vspace{-3mm}}
	\label{fig:noc_nbody16_32}
\end{figure}

Fig.~{\ref{fig:noc_nbody16_32}} compares N-body on \nocswc\ with BASE32 for increasing body numbers. The performance is measured in single-precision FLOPs/cycles. In Fig.~{\ref{fig:noc_nbody3}}, the performance is compared for 16 cores. It can be seen that the 16-core SMP system performs better for a smaller number of bodies ($<$2,048). 

In Fig.~{\ref{fig:noc_nbody4}}, a total of 32 cores are compared. As was the case for 16 cores, the SMP system performs better for smaller body counts. At body counts larger than 1,024, the performance of the SMP system drops significantly due to the fact that the working sets get larger than the available L1 data caches, resulting in increased bus pressure. The distributed computation, on the other hand, does not experience this bottleneck for the same number of bodies. 

In summary, it was shown that the benefit of employing a distributed system largely depends on the \textit{computation to communication ratio}. The naive Matmul does not benefit much from distributing the workload across multiple nodes; in fact, the performance worsens when more than eight nodes are used, due to the heavy communication requirements. The N-body simulation can profit from the NoC for large N, as the computation greatly dominates the communication.

Further, it was shown that the \textit{hybrid} approach, i.e., having multiple cores inside each node, can yield an additional performance gain, as intra-node parallelism can be exploited. The other benefit of hybrid systems is that less inter-node communication is required with the same total core count (e.g., (4,4) vs. (16,1)).

%% file: conclusion.tex
We presented a light-weight NoC for distributed memory MPSoC. Later, we presented ANDROMEDA, a RISC-V MPSoC exploration framework that incorporated the light-weight NoC. The generated RISC-V MPSoCs are prototyped on the Synopsys HAPS-80D Dual. The experimental evaluation demonstrated that the ANDROMEDA framework is easy to use for early-stage system prototyping. The user analyses of the benchmarks and applications help to identify the performance bottleneck in the NoC-based MPSoCs. In the future, we plan to extend support for other FPGA platforms and focus on RISC-V customization for further run-time reduction. 